\documentclass[letterpaper,titlepage,11pt]{article}
\usepackage{latexsym,amssymb,amstext,amsmath}
\usepackage{slashed}
\usepackage{mathrsfs}
\usepackage{color}
\usepackage{enumerate}
\usepackage{graphicx}
\usepackage{tikz}
\usepackage{etex}
\usepackage{latexsym}
\usepackage{cite}
\allowdisplaybreaks

\usepackage[
colorlinks=true,
linkcolor=blue,
urlcolor=red,
filecolor=green,
citecolor=red,
pdfstartview=FitV,
pdftitle={},
pdfauthor={Mehmet Ozkan},
pdfsubject={},
pdfkeywords={},
pdfpagemode=None,
bookmarksopen=true
]{hyperref}

\newcommand{\bea}{\setlength\arraycolsep{2pt} \begin{eqnarray}}
	\newcommand{\eea}{\end{eqnarray}}
\newcommand{\nn}{\nonumber}

\usepackage{hyperref}

\newsavebox{\uuunit}
\sbox{\uuunit}
{\setlength{\unitlength}{0.825em}
	\begin{picture}(0.6,0.7)
		\thinlines
		\put(0,0){\line(1,0){0.5}}
		\put(0.15,0){\line(0,1){0.7}}
		\put(0.35,0){\line(0,1){0.8}}
		\multiput(0.3,0.8)(-0.04,-0.02){12}{\rule{0.5pt}{0.5pt}}
		\end {picture}}

	\def\be{\begin{equation}}
		\def\ee{\end{equation}}
	\def\ba{\begin{array}}
		\def\ea{\end{array}}
	\def\bea{\begin{eqnarray}}
		\def\eea{\end{eqnarray}}
	\def\bd{\begin{displaymath}}
		\def\ed{\end{displaymath}}
	
	\def\nn{\nonumber}
	
	\def\nn{\nonumber}

	\makeatletter
	\@addtoreset{equation}{section}
	\makeatother
	
\begin{document}
		\begin{titlepage}
			
			\bigskip
			\begin{center}
				{\LARGE\bfseries  The Holographic c-theorem \\ [.15truecm] and Infinite-dimensional Lie Algebras}
				\\[10mm]
				\textbf{Eric A. Bergshoeff$^1$, Mehmet Ozkan$^2$ and  Mustafa Salih Z\"o\u{g}$^2$}\\[5mm]
				\vskip 25pt
				
				{\em  \hskip -.1truecm $^1$Van Swinderen Institute, University of Groningen, Nijenborgh 4, 9747 AG Groningen, The Netherlands  \vskip 10pt }
				
				{\em  \hskip -.1truecm $^2$Department of Physics, Istanbul Technical University,  \\
					Maslak 34469 Istanbul, Turkey  \vskip 10pt }

				{email: {\tt e.a.bergshoeff@rug.nl, ozkanmehm@itu.edu.tr, zog@itu.edu.tr}}
				
			\end{center}
			
			\vspace{3ex}

			\begin{center}
				{\bfseries Abstract}
			\end{center}
			\begin{quotation} \noindent
				
				We discuss a non-dynamical theory of gravity in three dimensions which is based on an infinite-dimensional Lie algebra that is closely related to an infinite-dimensional extended AdS algebra. We find an intriguing connection between on the one hand higher-derivative gravity theories that are consistent with the holographic c-theorem and on the other hand truncations of this infinite-dimensional Lie algebra that violate the Lie algebra structure. We show that in three dimensions different truncations reproduce, up to terms that do not contribute to the c-theorem, Chern-Simons-like gravity models describing extended 3D massive gravity theories. Performing the same procedure with similar truncations in dimensions larger than or equal to four reproduces higher derivative gravity models that are known in the literature to be consistent with the c-theorem but do not have an obvious connection to massive gravity like in three dimensions.
				
			\end{quotation}
			
			\vfill
			
			\flushleft{\today}
		\end{titlepage}
		\setcounter{page}{1}
		
		\newpage

		\section{Introduction}
		\paragraph{}
		One motivation to study higher-derivative corrections to the Einstein-Hilbert action is rooted in its connection to string theory where an infinite number of such corrections arise in the low-energy limit of string theory \cite{Candelas:1985en,Zwiebach:1985uq}. Although a priory there is a large number of higher-curvature terms that can be added to the Einstein-Hilbert action at each order of $\alpha^\prime$, many of those terms can be absorbed into a redefinition of the metric field \cite{Deser:1986xr}. Consequently, only terms that include a Riemann tensor actually contribute to the computation of physical quantities. In three dimensions, the Riemann tensor is not independent but can be expressed in terms of the metric, the Ricci scalar, and the Ricci tensor as the Weyl tensor identically vanishes. Thus, all higher-curvature corrections, if treated as perturbative interactions, can be absorbed into field redefinition and one is left with an action that only consists of an Einstein-Hilbert term, a cosmological constant, and a gravitational Chern-Simons term \cite{Gupta:2007th}.

Independent of string theory,  three-dimensional gravity theories with a {\sl finite} number of higher-derivative corrections have their own merit as modified theories of gravity. Three-dimensional general relativity with (or without) a cosmological constant can be written as a Chern-Simons gauge theory \cite{Achucarro:1987vz,Witten:1988hc} and has no dynamical degree of freedom. Higher-derivative terms then emerge as a mechanism to provide dynamics and lead to massive spin-2 modes in the spectrum \cite{Bergshoeff:2009hq} (see also \cite{Deser:1983tn,Nishino:2006cb}). There are various criteria to consider special combinations of higher-derivative terms. One criterion is to require the absence of a scalar ghost degree of freedom that arises for an arbitrary choice of the relative coefficients of the higher-derivative terms \cite{Boulware:1973my}. From a holographic perspective, demanding the existence of a holographic c-function is another way to single out particular combinations \cite{Sinha:2010ai,Paulos:2010ke,Gullu:2010st,Alkac:2018whk}.

Motivated by the Chern-Simons formulation of three-dimensional general relativity, one way to select special combinations of higher-derivative terms is to demand that they follow from what is called a Chern-Simons-like model of gravity that includes a  number of auxiliary Lorentz-vector valued one-form fields that, upon integrating out, yield the desired combination  \cite{Hohm:2012vh,Bergshoeff:2014bia,Afshar:2014ffa}. To be specific, consider the  Chern-Simons action
		\begin{eqnarray}
			S &=& \frac{k}{4\pi} \int {\rm{Tr}} \left( A \wedge dA + \frac23 A \wedge A \wedge A \right) \,,
		\end{eqnarray}
		where $k$ is the Chern-Simons coupling constant, the gauge field $A$ represents a Lie-algebra-valued one form and we take the trace using the non-degenerate invariant bilinear form of the Lie algebra. We  extend this action by replacing $A$ by a number of Lorentz-valued one-forms $A^{ra}$, where $a,b,\ldots$ are the Lorentz indices, and $r,s,t,\ldots$ are the flavor indices taking values in a $N$-dimensional \textit{flavor space}. The invariant bilinear form is then replaced by $\eta_{ab} g_{rs}$ where $g_{rs}$ is a symmetric, invertible  \textit{metric} on the $N$-dimensional flavour space. A Chern-simons like action is now obtained by replacing the structure constants of the Chern-Simons action  by $\epsilon_{abc} f_{rst}$ where the totally symmetric $f_{rst}$ is a \textit{flavor tensor} that does {\sl not} satisfy the Jacobi identity \cite{Bergshoeff:2014bia}. An $N$-flavor Chern-Simons-like theory consists of the Dreibein $e^a$, the spin-connection $\omega^a$, and $(N-2)$ auxiliary one-form fields. Following  \cite{Afshar:2014ffa}, the field equations describing what is called extended massive gravity are given by the following finite hierarchy of equations that can be solved for the auxiliary fields one after the other leaving the Dreibein $e^a$ as the only independent field:
		\begin{equation}
			\begin{split} \label{Hierarchy}
				& d e^a  + \epsilon^{abc} \omega_b e_c = 0\,, \\
				& R^a + \epsilon^{abc}e_b  h_{(1)c} = 0 \,, \\
				& d h_{(1)}^a + \epsilon^{abc} \omega_b h_{(1)c} +  \epsilon^{abc} e_b  f_{(1)c} = 0\,, \\
				& d f_{(1)}^a +  \epsilon^{abc} \omega_b f_{(1)c} + \frac12\epsilon^{abc} f_{(1)b} f_{(1)c} + \epsilon^{abc}e_b  h_{(2)c} = 0\,,\\
				&  \qquad \quad \vdots \\
			\end{split}
		\end{equation}
		where $R^a$ is the dualized curvature two-form
		\begin{equation}
			R^a= d\omega^a + \frac12 \epsilon^{abc}  \omega_b \omega_c \,,
		\end{equation}
		and $\omega^a, f^a_{(n)}$ and  $h^a_{(n)}\ (n=N/2-1)$ refer to the auxiliary fields. Assuming that the Dreibein $e^a$ is invertible, the first equation determines $\omega^a= \omega^a(e)$ to be the torsionless spin-connection. Subsequently, the second equation is used to solve for $h_{(1)}^a$ in terms of $e^a$ and, next, the third equation is used to solve for $f_{(1)}^a$ in terms of $e^a$ etc.  The remaining equations can be solved to determine all auxiliary fields, $f_{(n)}^a$ and  $h_{(n)}^a$, in terms of $e^a$ and its derivatives. Assuming a finite set of equations,  the last equation determines the dynamical equation of motion of the theory under consideration. This particular method of constructing higher derivative gravity models uses the Schouten and  Cotton tensors as the basic building blocks. This is due to the fact that, in the process of solving the hierarchy of equations \eqref{Hierarchy}, one finds that the lowest order auxiliary fields, $h_{(1)}$ and $f_{(1)}$, are given by the Schouten and  Cotton tensor, respectively.

Another guideline to extend the Einstein-Hilbert action with higher curvature terms is the existence of a holographic c-function \cite{Myers:2010xs}. For the formulation of the holographic c-theorem in Einstein gravity, one begins with the following Ansatz describing the renormalization group flow between different critical points \cite{Freedman:1999gp,Myers:2010xs}:
			\begin{eqnarray} \label{BackgroundC}
				ds^2 = 2 e^{2\mathcal A(r)} (-dt^2 + d\textbf{x}^2) + dr^2
			\end{eqnarray}
with 	$\mathcal A(r)$ an arbitrary function of $r$.	Requiring this Ansatz to be a solution of the Einstein equations, implies
\begin{equation}
{\mathcal A}^{\prime\prime}(r) = T^t{}_t - T^r{}_r \leq 0\,,
 \end{equation}
 where $T_{ab}$ is the stress-energy tensor of the matter sector which is assumed to satisfy the null energy condition. This inequality allows one to define a function $c(r)$
			\begin{eqnarray}
				c(r) = \frac{1}{\ell_P \mathcal A^\prime (r)}\,,
			\end{eqnarray}
			where $\ell_P$ is the Planck length and which, due to the null energy condition,  is monotonically increasing.
		
Extending the  gravitational sector with  higher-derivative terms and assuming that the matter part does not include mixed, higher derivative interactions, the necessary condition to define  a monotonically increasing function $c(r)$ is that the equations of motion remain second derivative when expanded on the background \eqref{BackgroundC} \cite{Myers:2010tj, Liu:2010xc,Oliva:2010eb}.  For $D \geq 4$, this condition naturally selects Lovelock theories \cite{Lanczos:1938sf,Lovelock:1971yv,Lovelock:1972vz} as they have second-order field equations for any choice of metric. Given the fact that the Weyl tensor and Cotton tensor vanishes on the background \eqref{BackgroundC} and the fact that the Riemann tensor  decomposes into the sum of a Weyl tensor and  a Schouten tensor, one can show that the $D$-dimensional Lagrangian that is compatible with the holographic c-theorem is given by \cite{Paulos:2012xe}
			\begin{eqnarray}\label{CTheorem}
				e^{-1} {\mathcal L} &=& R  - 2 \Lambda  +  \sum_{n=0}^D \alpha_n \mathcal{P}^{(n)} (S_a{}^b ) \,,
			\end{eqnarray}
			where the $\alpha_n$ are free parameters, $S_{ab}$ is the $D$-dimensional Schouten tensor, and the polynomial $\mathcal{P}^{(n)} (S_a{}^b )$ is given by
			\begin{eqnarray}
				\mathcal{P}^{(n)} (S_a{}^b ) = \delta_{[b_1 \ldots b_n]}^{a_1 \ldots a_n} \,  S_{a_1}{}^{b_1} \cdots S_{a_n}{}^{b_n} \,.
			\end{eqnarray}

		At first sight, this form of the Lagrangian seems to limit the existence of compatible models to a certain order in the Schouten tensor depending on the dimension of spacetime. However,  one can define a \textit{holographic limit} by first taking a $D > d$ dimensional model, followed by the rescaling of the Lagrangian by $1/(D-d)$ and next taking the limit $D \to d$ \cite{Alkac:2020zhg}. As an example, let us consider a $D$-dimensional model with $D>3$. The $\mathcal{P}^{(4)} (S_a{}^b)$ contribution, which vanishes for $D=3$ due to the antisymmetrization of four indices, is given by
			\begin{eqnarray} \label{1m6Contribution}
				(D-3)\left( - \frac{(D-2)}{D (D-1)} (\slashed S_{\mu\nu}^2)^2 + \frac{8}{D} \slashed S_{\mu\nu}^3 S - \frac{6  (D-2)}{D^2} S^2 \slashed S_{\mu\nu}^2 - \frac{(D^2 - 3D + 2)}{D^3} S^4 \right) \,,
			\end{eqnarray}
			where we decompose the Schouten tensor into its traceless part $(\slashed S_{\mu\nu})$ and the trace part $(S)$ and where we have used the following identity for the traceless Schouten tensor \cite{Alkac:2020zhg}
			\begin{eqnarray}
				D (D-1) \slashed S_{\mu\nu}^4 = (D^2 - 3 D + 3) (\slashed S_{\mu\nu}^2)^2  \,.
			\end{eqnarray}
			Rescaling the Lagrangian \eqref{1m6Contribution} with a factor  $1/(D-3)$ followed by taking the limit $D \to 3$ precisely recovers the $1/m^6$ contribution to the holographic c-theorem in three-dimensions \cite{Sinha:2010ai}. Thus, there is an infinite number of higher-derivative gravity models in any dimension and at any order that is compatible with the holographic c-theorem with the  Schouten tensor as the basic building block. Remarkably, this same mechanism also reproduce the three-dimensional massive gravity models  \cite{Sinha:2010ai,Paulos:2010ke,Gullu:2010pc} whose Chern-Simons-like formulation coincides with the hierarchy of equations \eqref{Hierarchy} up to terms that do not contribute to the c-function \cite{Afshar:2014ffa}.

		A characteristic feature of the Chern-Simons formulation of three-dimensional gravity is that it does not describe any propagating degree of freedom and that the equations of motion are given in terms of group-theoretical curvatures\cite{Achucarro:1987vz,Witten:1988hc} with an underlying Lie algebra. If the hierarchy of field equations \eqref{Hierarchy} for the $N$ flavor model is extended to an {\sl infinite} number of flavors, it provides another example of a three-dimensional theory of gravity with no dynamics. The reason is that in the infinite-dimensional case, there is no last equation representing any dynamics. This curious property of \eqref{Hierarchy} leads one to consider the possibility that there is an infinite-dimensional Lie algebra that underlies the infinite number of field equations such they can all be formulated in terms of the group-theoretical curvatures corresponding to this infinite-dimensional Lie algebra. In this work, we will show that this is indeed the case.  Furthermore, we will show that the truncations of this infinite-dimensional Lie algebra to a {\sl finite} number of generators violates the Lie algebra structure, i.e.~the Jacobi identities are not satisfied after the truncation. As a result, after truncation, the resulting actions become {\sl Chern-Simons-like} and describe extended massive gravity models up to terms that do not contribute to the c-function. We will show that this infinite-dimensional Lie algebra is closely related to an infinite-dimensional extended AdS$_3$ algebra, The latter algebra can be truncated consistently but these truncations lead to Chern-Simons actions that do not describe massive gravity.
		
		Having established a relation between the holographic c-theorem and (truncations of) an infinite-dimensional Lie algebra in 3D, we address the $D>3$ case by following the same procedure. A key ingredient in this generalized procedure is that we use a first-order formulation of gravity that avoids the use of inverse Vielbein fields and that we replace the infinite-dimensional AdS$_3$ algebra by an infinite-dimensional AdS$_{\rm D}$ algebra. In doing so, we find that the connection between the holographic c-theorem and higher-derivative gravity models remains but that there is no longer a relation with higher-dimensional massive gravity.

		\section{The infinite-dimensional Lie algebra underlying the 3D holographic c-theorem}\label{3DMassiveGravity}
		\paragraph{}
		The infinite-dimensional Lie algebra that we claim to underlie the 3D holographic c-theorem has two sets of generators, $P_a^{(n)}$ and $J_a^{(n)}$, with  $n \geq 0$, whose non-zero commutation relations are given by
		\begin{align} \label{Algebra}
			[{J}_a^{(m)}, {J}_b^{(n)}] &= \epsilon_{abc}{J}^{c\, (m+n)} \,, &	[{J}_a^{(m)}, {P}_b^{(n)}] &=  \epsilon_{abc}{P}^{c\, (m+n)}\,,\nonumber\\
			[{P}_a^{(m)}, {P}_b^{(n)}] &= -\mu^2  \epsilon_{abc}{J}^{c\, (m+n-1)}\,,
		\end{align}
		where $\mu$ is an arbitrary parameter of mass dimension $1$. Using that $J_a^{(-1)}=0$, the generators  $J_a \equiv J_a^{(0)}$ and $P_a \equiv P_a^{(0)}$ are the standard generators of Lorentz transformations and translations, respectively. The algebra \eqref{Algebra} has several noteworthy properties. When $\mu = 0$,
the infinite-dimensional Lie algebra can be consistently truncated to any finite number of generators  $P_a^{(n)}\,,J_a^{(n)}$ with  $n \leq N$,
Thus, a Chern-Simons action can be formed by assigning a gauge field to each generator and using the invariant bilinear form of the truncated finite-dimensional Lie algebra. In that case, the field equations do not describe the propagating degrees of freedom of a massive gravity model but instead, they correspond to the coupling of a set of gauge fields, $\{h^a_{(n)}, f^a_{(n)}\}$, to gravity.

The situation is rather different when $\mu \neq 0$. In that case, the closure of the algebra is satisfied only if it is infinite-dimensional or if we take $N=0$. To see that, let us consider a truncation of the algebra \eqref{Algebra} at the order $2N-1\ (N\ne 0)$, i.e.~the highest order generators are $P_a^{2N-1}$ and $J^{2N-1}_a$. The triplet of generators that fails to satisfy the Jacobi identity is given by  $(P_a, P_b^N, J_c^N)$. This is due to the fact that the $[P_b^N, J_c^N]$ commutator vanishes according to \eqref{Algebra}. However, the contribution from this commutator is essential to satisfy the Jacobi identity for the triplet $(P_a, P_b^N, J_c^N)$. Clearly, this particular Jacobi identity is satisfied in the infinite-dimensional case.

Making the following assignment of gauge fields to each generator
		\begin{eqnarray}\label{Curvatures}
			A = e^a P_a + \omega^a J_a + h^a_{(n)} P_{a}^{(n+1)}  + f^a_{(n)} J_{a}^{(n+1)}  \,,
		\end{eqnarray}
		we can calculate the group-theoretical curvatures which, when set to zero,  are the field equations of the Chern-Simons theory. This leads to an infinite set of equations that can be conveniently summarized by the following set of equations containing a parameter $\lambda$:
		\begin{eqnarray}
			\label{Set1}
			0 &=& \left( de^a + \epsilon^{abc} \omega_b  e_c \right) +   \sum_{n=1}^\infty \lambda^{2n}\left[ Dh^a_{(n)}  + \epsilon^{abc} e_b  f_{(n)c} \right] +   \sum_{n,m=1}^{\infty}  \lambda^{2(n+m)}  \epsilon^{abc} h_{(n)b}  f_{(m)c} \,, \\
			0 &=&  \left( R^a  - \mu^2  \epsilon^{abc} e _b  h_{1c}  \right) + \frac12 \sum_{n,m=1}^{\infty}  \lambda^{2(n+m)}  \epsilon^{abc}  f_{(n)b}  f_{(m)c}  \nonumber\\
			&&  + \sum_{n=1}^{\infty} \lambda^{2n}\left[  D f_{(n)}^a  - \mu^2  \epsilon^{abc}  e_b  h_{(n+1)c} \right] -  \frac12 \mu^2 \sum_{n,m=1}^{\infty}  \lambda^{2(n+m-1)} \epsilon^{abc}  h_{(n)b}  h_{(m)c} \,. \label{Set2}
		\end{eqnarray}
		Here $D$ represents the standard Lorentz covariant derivative, i.e.
		\begin{eqnarray}
			Df^a &=& df^a + \epsilon^{abc} \omega_b f_c \,.
		\end{eqnarray}
		
\begin{figure}
			\centering
			\includegraphics[scale=0.4]{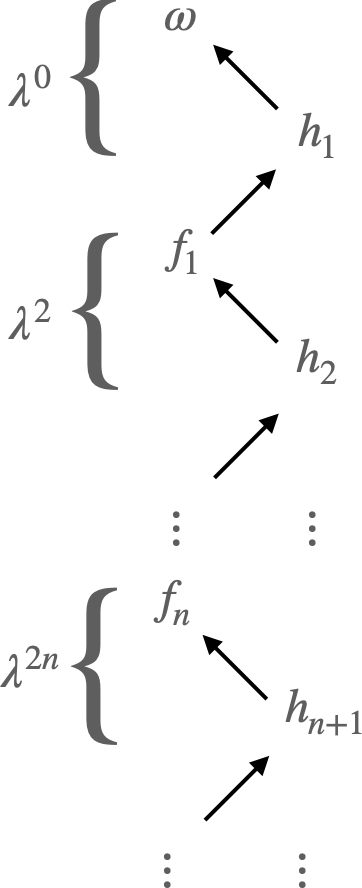}
			\caption{\footnotesize The hierarchy of equations according to eqs.~\eqref{Set1} and \eqref{Set2}. One needs to determine the field at the head of the arrow in order to solve for the field at the tail of the same arrow. At $\lambda^0$ order, the equations determine $\omega$ as the torsionless spin connection and $h_1$ to be the Schouten tensor. Then, for each following order at $\lambda^{2n}$, the remaining auxiliary fields are solved in terms of the dualized Dreibein curvature two-form $R^a$.}
			\label{HierarchyFig}
		\end{figure}

		The different equations of motion or curvature constraints are found by setting all terms multiplying a given order of $\lambda$ equal to zero.
	The first set of equations, given by (\ref{Set1}), represent all $R(P)$ curvature constraints. Similarly, the second set, given by (\ref{Set2}), represent all $R(J)$ curvature constraints.  Assuming that $e^a$ is the invertible Dreibein, the equation  multiplying $\lambda^0$ in the first set implies that $\omega^a$ is the torsionless spin-connection $\omega^a= \omega^a(e)$. Subsequently, since $e^a$ is invertible, the equation multiplying $\lambda^0$ in the second set determines $h^a_{(1)}$ to be the Schouten tensor. Next, the $\lambda^2$ terms in the first set can be used to solve $f_{(1)}^a$ in terms of $h_{(1)}^{a}$ and the  $\lambda^2$ terms in the second set can be used to solve $h_{(2)}^a$ in terms of $f_{(1)}^{a}$ etc.  Hence, the combined set of equations gives rise to the desired hierarchy of equations \eqref{Hierarchy} for the infinite flavor case, see Fig.\ref{HierarchyFig}. Note that in the infinite flavor case, there is no last equation representing any dynamics which is as expected for a Chern-Simons theory.
		
		Now that we have shown that the infinite-dimensional Lie algebra \eqref{Algebra} gives rise to a soluble set of equations of the form \eqref{Hierarchy}, we may discuss the inconsistent truncations and the resulting Chern-Simons-like models of gravity. If the algebra, hence the field equations, is truncated at an order $\lambda^{2N}$, then the hierarchy continues up to that order. There are two possibilities. If the highest field in the truncation is $f_{(N)}$, then the dynamical field equation of the theory includes a $Df_{(N)}$ term. Assigning the following parity-properties
		\begin{eqnarray}
			\text{Even}: \, \{e^a, h_{(n)}^a\}\,, \quad 	\text{Odd}: \, \{\omega^a, f_{(n)}^a\} \,,\quad n \leq N\,,
		\end{eqnarray}
		this implies that such a truncation gives rise to parity-even massive-gravity models, i.e.~the action giving rise to this equation of motion is parity-even. Similarly, if $h_{(N)}$ is the highest field in the truncation, the dynamical field equation of the theory includes $Dh_{(N)}$ and  hence the theory is parity-odd.
		
		Let us now see how this works by working out a few examples. The simplest case is to consider a truncation to only $e^a$ and $\omega^a$ and the corresponding generators $P_a$ and $J_a$, respectively. In that case, we are left with  the Poincar\'e algebra as a truncation of (\ref{Algebra})
		\begin{eqnarray}
			[J_a, J_b] = \epsilon_{abc} J^c \,, \qquad [J_a, P_b] = \epsilon_{abc} P^c  \,,
		\end{eqnarray}
		and the corresponding field equations are the torsion-free connection equation and the Einstein equation. Note that this is a consistent truncation  and therefore leads to a Chern-Simons theory.  According to the hierarchy, Fig.\ref{HierarchyFig}, we next include the field $h_1$ and the corresponding generator $P_{(1)}^a$. Thus, we add the following additional former structure constants which we now interpret as the \textit{flavor tensors} of the Chern-Simons-like theory
		\begin{equation}
			[J_a, H_b] = \epsilon_{abc} H^c\,, \qquad [P_a, H_b] = - \mu^2 \epsilon_{abc} J^c\,,
		\end{equation}	
		where we have set $P_a^{(1)} \equiv H_a$. Following (\ref{Set1}) and (\ref{Set2}), the field equations then read
		\begin{eqnarray}
			0 &=& de^a +  \epsilon^{abc} \omega_b e_c \,,\nonumber\\
			0 &=& R^a - \mu^2 \epsilon^{abc} e_b h_{(1)c}\,,\nonumber\\
			0 &=&  Dh^a_{(1)} \,.
		\end{eqnarray}
		These are the field equations of 3D conformal gravity  as given in \cite{Afshar:2014ffa} up  to a rescaling of $h^a_{(1)} \to h^a_{(1)} /\mu^2$. This  theory is parity-odd as expected.

We can continue in accordance with Fig.\ref{HierarchyFig} and add a new field $f_1$ and the corresponding generator $J_{(1)}^a$. According to \eqref{Algebra}, this gives us three more flavor tensors
		\begin{align}
			[J_a,F_b] & = \epsilon_{abc} F^c \,,  & [F_a, P_b] & = \epsilon_{abc} H^c\,, &	[H_a, H_b] & = -\mu^2 \epsilon_{abc} F^c \,,
		\end{align}
		where we have set $J^{(1)}_a \equiv F_a$. The field equations based on these flavor tensors follow from (\ref{Set1}), (\ref{Set2}) and read
		\begin{align}
			0 &= de^a +  \epsilon^{abc} \omega_b e_c\,,\nonumber\\
			0 &= R^a - \mu^2 \epsilon^{abc} e_b h_{(1)c} \,,\nonumber\\
			0 &= Dh^a_{(1)} + \epsilon^{abc} e_b f_{(1)c} \,,\nonumber\\
			0 &= Df^a_{(1)} - \frac12 \mu^2 \epsilon^{abc} h_{(1)b} h_{(1)c} \,.
		\end{align}
		These are precisely the field equations for the curvature-squared part of new massive gravity \cite{Deser:2009hb,Afshar:2014ffa,Bergshoeff:2014bia}, which is parity even. It is a straightforward exercise to show that the next order models are the Chern-Simons-like formulation of the five-derivative part of 3D extended conformal gravity involving the set $\{e, \omega, h_1, f_1, h_1\}$ and the six-derivative part of 3D extended new massive gravity involving  the set $\{e, \omega, h_1, f_1,h_2, f_2\}$ \cite{Afshar:2014ffa,Sinha:2010ai}. All the gravity models discussed above precisely give rise to the curvature combinations predicted by the c-theorem up to curvature terms that do not contribute to the c-function \cite{Afshar:2014ffa,Sinha:2010ai}. Note, however, that beyond the six-derivative order the holographic c-theorem alone cannot fix all relative coefficients, thus the compatible higher-curvature models include free parameters \cite{Paulos:2010ke}. In that case, the Chern-Simons-like models, which we obtain by the truncation of the infinite-dimensional algebra \eqref{Algebra}, are still compatible, but they correspond to a subclass of compatible theories whose spectrum is free from the scalar ghost \cite{Afshar:2014ffa}. 
		
		For a given truncation, we may integrate the field equations \eqref{Set1} and \eqref{Set2} to an action. We should distinguish between parity-even and parity-odd truncations. The action for an $N$-th order parity-even Chern-Simons-like model, with $f_{(N)}^a$ as highest-order field, is given by
		\begin{align}
			L_N^{\rm even} &= f_{(N)}^a  De_a + h_{(N)}^a R_a  - \frac12 \mu^2 \epsilon^{abc} e_a h_{(1)b}  h_{(N)c}  + \sum_{n=1}^{N-1}   \left[ f_{(N-n)}^a   Dh_{(n)a} + \frac12 \epsilon^{abc} e_a f_{(N-n)b} f_{(n)c} \right]   \nonumber\\
			&- \frac12 \mu^2 \sum_{n=1}^{N} \epsilon^{abc} e_a  h_{(n+1)b}  h_{(N-n)c}   +\frac12 \sum_{m,n=1}^{N}  \epsilon^{abc} h_{(m)b}  f_{(n)c}  f_{(N- n-m)c}  \nonumber\\
			&- \frac16 \mu^2 \sum_{m,n=1}^{N} \epsilon^{abc} h_{(m)a}  h_{(n)b}  h_{(N- n- m+1)c} \,.
			\label{ParityEven}
		\end{align}
		Since $N \geq 1$, the Lagrangian \eqref{ParityEven} does not give the part of the Dreibein equation of motion that follows from the Einstein-Hilbert and cosmological term. However, we may consistently add these terms to the above Lagrangian since it does not effect the other equations of motion that are used to solve for $f^a_{(n)}$, $h^a_{(n)}$ and $\omega$ in terms of $e$ and its derivatives. Upon adding these terms,  the full dynamical Dreibein equation follows from the following modified Lagrangian:
		\begin{eqnarray}
			L_N^{\prime\, {\rm even}} = \sigma e^a  R_a + \frac16 \mu^2 \epsilon^{abc} e_a  e_b  e_c + \sum_{N \geq 1} a_N L^{\rm even}_N \,,
			\label{EvenBI}
		\end{eqnarray}
		where $\sigma = 0, \pm1$ and $a_N$ are dimensionless constants.

Similarly, the action for an $N$-th order parity-odd Chern-Simons-like model, with $h_{(N)}^a$ as highest-order field, is given by
		\begin{eqnarray}\label{ParityOdd}
			L_N^{\rm odd} &=& f_{(N-1)}^a R_a -\mu^2 e_a Dh_{(N)}^a -\frac{\mu^2}{2} \epsilon^{abc} e_a e_b f_{(N-1)c} + \frac{1}{2}\sum_{n=1}^{N-1}  f_{(n)}^a Df_{(N-n-1)a}  \nonumber\\
			&& -\frac{\mu^2}{2}\sum_{n=1}^{N} h_{(n)}^a  Dh_{(N-n)a}  -\mu^2 \sum_{n=1}^{N-1}  \epsilon^{abc} e_a h_{(n)b} f_{(N-n)c} \nonumber\\
			&& - \frac{\mu^2}{2}\sum_{n,m=1}^{N}  \epsilon^{abc} f_{(n)a} h_{(m)b} h_{(N-n-m)c}  +\frac{1}{6}\sum_{n,m=1}^{N} \epsilon^{abc}  f_{(n)a}  f_{(m)b}  f_{(N-n-m-1)c} \,.
		\end{eqnarray}
		Here, we restrict to $N>1$ and  hence the Lagrangian \eqref{ParityOdd}   does not contain the zeroth-order 3D conformal gravity Lagrangian. Adding this Lagrangian by hand, we obtain the following modified Lagrangian:
		\begin{eqnarray}
			L_N^{\prime\, {\rm odd}} &=& \frac{1}{2} \ell \left(\omega^a d\omega_a + \frac13 \epsilon^{abc} \omega_a \omega_b \omega_c \right) - \frac{1}{2\ell} e^a De_a  - \frac{1}{\ell} e_a Dh_1^a + \sum_{N>1}  a_N L_N^{\rm odd} \,.
		\end{eqnarray}
		
This concludes our discussion of the 3D infinite-dimensional Lie algebra underlying the 3D holographic c-theorem.
		
		\section{Relation with the  infinite-dimensional extended AdS$_3$ Lie algebra}\label{RelatedAlgebra}
		\paragraph{}
		
		It turns out that the infinite-dimensional Lie algebra \eqref{Algebra} underlying the 3D c-theorem is closely related to another infinite-dimensional Lie algebra that, in contrast with the algebra \eqref{Algebra},  can be consistently truncated to each finite order.  This infinite-dimensional Lie algebra is obtained by making a  Lie algebra expansion \cite{Hatsuda:2001pp,deAzcarraga:2002xi,deAzcarraga:2007et} of the AdS$_3$ algebra
		\begin{align} \label{AdSAlgebra}
			[{J}_a, {J}_b] &= \epsilon_{abc}{J}^{c} \,,\quad &	[{J}_a, {P}_b] &=  \epsilon_{abc}{P}^{c}\,, & [{P}_a, {P}_b]   & = -\Lambda \epsilon_{abc}{J}^{c}\,,
		\end{align}
		where $\Lambda$ is a mass parameter representing the cosmological constant.
		To perform such a an expansion we should write AdS$_3 = V_0 \oplus V_1$ where $V_0$ and $V_1$ are two subspaces with
		\begin{eqnarray}
			[V_0, V_0] \subset V_0 \,, \quad [V_0, V_1] \subset V_1 \,,\quad [V_1, V_1] \subset V_0 \,.
		\end{eqnarray}
		The subspace $V_0$ ($V_1$) then contains the generators that are expanded in terms of even (odd)  powers of an expansion parameter $\lambda$. We consider the special case that $V_0$ = AdS$_3$ and $V_1=0$.
		Upon expansion, this leads to the infinite-dimensional extended AdS$_3$ algebra
		\begin{align}\label{CoAdjoint}
			[{J}_a^{(m)}, {J}_b^{(n)}] &= \epsilon_{abc}{J}^{c\, (m+n)}\,,  &	[{J}_a^{(m)}, {P}_b^{(n)}] &= \epsilon_{abc}{P}^{c\, (m+n)}\,, &	[{P}_a^{(m)}, {P}_b^{(n)}] &= -\Lambda \epsilon_{abc}{J}^{c(m+n)}\,.
		\end{align}
		Associating a gauge field to each generator,
		\begin{eqnarray}
			A = E^a P_a + \Omega^a J_a \,,
		\end{eqnarray}
		we may expand these gauge fields in even powers of $\lambda$   as
		\begin{eqnarray} \label{expansion}
			E^a = e^a + \sum_{n=1}^{\infty} \lambda^{2n} h_{(n)}^a \,, \quad \Omega^{a} = \omega^a + \sum_{n=1}^{\infty} \lambda^{2n} f_{(n)}^a  \,.\quad
		\end{eqnarray}
		We may use the expansion (\ref{expansion}) of the fields  in the group-theoretical curvatures of $E^a$ and $\Omega^a$ and expand the field equations of the cosmological (or exotic \cite{Witten:1988hc,Townsend:2013ela}) Einstein-Hilbert action
		\begin{eqnarray}\label{ExpE}
			0 &=& dE^a + \epsilon^{abc} \Omega_b E_c \,, \\
			0 &=& d\Omega^a + \frac12 \epsilon^{abc} \Omega_ b \Omega_c - \frac12 \Lambda \epsilon^{abc} E_b E_c \,.
			\label{ExpO}
		\end{eqnarray}
		The expansion of (\ref{ExpE}) precisely generates the first set of equations (\ref{Set1}), providing a set of soluble equations, i.e.~they determine $\omega^a$ to be the torsionless spin-connection $\omega^a(e)$ and all the $f^a_{(n)}$'s can be solved once the $h^a_{(n)}$'s are known, leaving $e^a$ to be the only fundamental object. However,  the expansion of (\ref{ExpO}), does not provide a soluble set, like the second set of equations (\ref{Set2}). Instead,  it leads to the following set of equations:
		\begin{eqnarray}
			0 &=& \left( R^a  - \frac12 \Lambda \epsilon^{abc} e_b e_c  \right)   + \sum_{n=1}^{N} \lambda^{2n}\left[  Df_{(n)} - \Lambda  \epsilon^{abc} e_b  h_{(n)c} \right] + \ldots \,.
			\label{OExp}
		\end{eqnarray}
		Here, the ellipses refer to terms quadratic in the fields that are not relevant to our discussion, since they do not contain an invertible Dreibein.  They can be calculated  by implementing the  expansion \eqref{expansion} in eq.~\eqref{ExpO}. Comparing the infinite-dimensional Lie algebras \eqref{Algebra} and \eqref{CoAdjoint}, we see that they only differ in the $[P,P]$ commutator:
		\begin{align}\label{shift}
			\eqref{Algebra}:  [{P}_a^{(m)}, {P}_b^{(n)}] = -\mu^2 \epsilon_{abc}{J}^{c\, (m+n-1)}
			\ \ \ \Leftrightarrow\ \ \
			\eqref{CoAdjoint}:  [{P}_a^{(m)}, {P}_b^{(n)}] = -\Lambda \epsilon_{abc}{J}^{c(m+n)}\,.
		\end{align}
		This has the important consequence that in both cases the expansion of the second set of equations  starts in a different way:
		\begin{equation} \label{compare}
			\left( R^a  - \mu^2  \epsilon^{abc} e _b  h_{(1)c}  \right)
			\ \ \ \Leftrightarrow\ \ \
			\left( R^a  - \frac12 \Lambda \epsilon^{abc} e_b e_c  \right)\,.
		\end{equation}
		It is clear that in the first case we can use the lowest order equation to solve for  $h^a_{(1)}$ in terms of the Schouten tensor whereas in the second case we obtain the Einstein equation. Only in the first case we obtain, together with  the expansion of (\ref{ExpE}) that leads  to the first set of equations \eqref{Set1}, a completely soluble set of equations. The corresponding Lagrangian is Chern-Simons like, describes extended massive gravity and is consistent with the c-theorem. On the contrary, in the second case, each consistent finite truncation leads to a Chern-Simons model describing the coupling of a set of gauge fields to gravity. This infinite-dimensional extended AdS$_3$ algebra will be used in the next section as a guide to derive the infinite-dimensional Lie algebra underlying the c-theorem in $D\geq 4$ dimensions.

We note that the set of soluble field equations \eqref{Hierarchy} and those corresponding to the infinite-dimensional extended AdS$_3$ algebra are formally related to each other by replacing the cosmological constant $\Lambda$ used in the AdS case by the massive parameter $\mu$ used in the other case multiplied by the inverse square of the expansion parameter $\lambda$ as follows:
\begin{equation}\label{relation}
\Lambda = \mu^2/\lambda^2\,.
\end{equation}
		
		It is a unique property of three-dimensions that we were able to integrate the $N$-th order field equations to an $N$-th order Chern-Simons like Lagrangian describing extended massive gravity. The reason for that is because in three-dimensions, the field equations of a Chern-Simons theory are given in terms of  the group-theoretical curvatures. This property no longer  holds in $D>3$ dimensions neither does the relation with massive gravity. In fact, a proposal for a 4D new massive gravity theory as in \cite{Bergshoeff:2012ud}  involves the use of mixed-symmetry potentials with no obvious relation to an algebra.

			Before ending this section, let us briefly comment on our assumption that the matter part does not include mixed, higher derivative interactions when implementing the holographic c-theorem in higher-derivative gravity models. In principle, it is tempting to start by a cosmological Einstein-Hilbert action with a minimally coupled matter sector that satisfies the null energy condition and expands all fields, including the matter fields, in even powers of $\lambda$ and rescale the cosmological constant in accordance with \eqref{relation}. However, unlike Chern-Simons theory, the minimal matter coupling would include an explicit metric field, and the expansion procedure yield non-trivial coupling between the gravity sector and the matter sector. To evade this problem, we only expand the gravity sector and include the minimally coupled matter fields that satisfy the null energy condition once the expansion procedure is complete. Note that for parity-even models, which are the ones that contribute non-trivially to holographic c-theorem, the dynamical field equation for the gravity sector always arise as the vielbein equation. Hence the inclusion of the minimally coupled matter sector does not spoil the solubility of the hierarchy of equations.
		
		\section{Extension to $D \geq 4$ dimensions}
		\paragraph{}
			It is natural to generalize our previous observations to higher dimensions and consider whether the infinite-dimensional structures we find in 3D also work in higher dimensions. It is known that there exist infinitely many models that are compatible with the holographic c-theorem in any dimensions where the basic building block is the Schouten tensor $S_{\mu\nu}$ \cite{Paulos:2012xe}. The Lagrangian describing these models is of the generic form \eqref{CTheorem}. On the other hand, the soluble set of field equations that we considered, which are compatible with an underlying infinite-dimensional algebra, uses the Schouten $(S_{\mu\nu})$ and the Cotton $(C_{\mu\nu\rho})$ tensors as basic building blocks:
			\begin{align}\label{SchoutenCotton}
				S_{\mu\nu} = \frac{1}{(D-2)}  \left( R_{\mu\nu} - \frac1{2(D-1)} g_{\mu\nu} R \right)\,,  \qquad  C_{\mu\nu\rho}  = (D-2) \left( \nabla_\lambda S_{\mu\nu} - \nabla_\nu S_{\mu\lambda} \right) \,.
			\end{align}
		
To see the connection between the set of soluble equations and the holographic c-theorem, let us focus on $D=4$ as $D > 4$ is a simple generalization. The analogue of the hierarchy of equations \eqref{Hierarchy} that gives rise to models that are compatible with the holographic c-theorem takes the following form in four dimensions:
		\begin{equation}
			\begin{split} \label{Hierarchy4D}
				& e \wedge De = 0\,, \\
				& e \wedge R(\omega)  - \frac12 \mu^2 e \wedge e  \wedge  h_{(1)} = 0 \,, \\
				& e \wedge D h_{(1)} + e \wedge e \wedge  f_{(1)} = 0\,, \\
				&  \qquad \quad \vdots \\
			\end{split}
		\end{equation}
		For simplicity, we have refrained from denoting the explicit Lorentz indices of the different fields. As in three dimensions, this set of field equations can be derived from the field equations of the four-dimensional Einstein-Cartan action by means of an expansion of the gauge fields and the rescaling of the cosmological constant. However, unlike the three dimensions, these field equations pick up extra Vierbein factors. These factors provide the necessary trace decompositions and allow us to determine $\omega^{ab}$ to be the torsionless spin connection and define the auxiliary fields, $h_{(n)}^a$ and $f_{(n)}^{ab}$, in terms of the Schouten tensor, the Cotton tensor, and its descendants.  When the number of auxiliary fields $\{h_{(n)}^a, f_{(n)}^{ab}\}$ is taken to be infinite, these field equations become a set of equations in terms of curvatures that transform covariantly under the following infinite-dimensional algebra
\begin{align}\label{4dExpandedAlgebra}
	[P_a^{(m)}, M_{bc}^{(n)}] &= \eta_{a[b} P_{c]}^{(m+n)}\,, & [M_{ab}^{(m)}, M_{cd}^{(n)}] &= 4 \eta_{[b[c} M_{a]d]}^{(m+n)}\,, \nn\\
	[P_a^{(m)},P_b^{(n)} ] &=  - \frac14 \mu^2 M_{ab}^{(m+n-1)} \,.
\end{align}
This algebra is the four-dimensional generalization of \eqref{Algebra}. One may verify that starting with the field equations of the first order formulation of cosmological Einstein action and expanding them by using the standard Lie algebra expansion with $V_0$ = AdS$_4$ and $V_1=0$, followed by a rescaling of the cosmological constant by $\lambda^{-2}$ as we did in Section \ref{RelatedAlgebra},  produces an infinite number of field equations which transform covariantly under the infinite-dimensional Lie algebra \eqref{4dExpandedAlgebra}.

To perform the Lie algebra expansion at the level of Lagrangian, it is convenient to use the following first-order form of the  cosmological Einstein-Hilbert action that does not contain inverse Vierbeine:
		\begin{align}
			L = \frac12 E\wedge E \wedge R(\Omega) - \frac{1}{24}  \Lambda\, E \wedge E \wedge E \wedge E \,.
		\end{align}
		We have defined here the curvature two-form as
		\begin{align}
			R^{ab}(\Omega) = \frac12 R^{ab}{}_{cd} (\Omega) E^c \wedge E^d \,.
		\end{align}
		This action may be expanded using the following expansion rules
		\begin{align}
			E^a = e^a + \sum_{n=1}^N \lambda^{2n} h_{(n)}^a \,, \quad \Omega^{ab} = \omega^{ab} + \sum_{n=1}^N \lambda^{2n} f_{(n)}^{ab} \,.
		\end{align}
		To obtain  the  Lagrangian corresponding to the soluble models we need to rescale the cosmological constant like we did in the 3D case, see eq.~\eqref{relation}.  In this case, after truncation  the would-be structure constants become flavour tensors.
		
		It is instructive to  work out a few models to see the resulting Lagrangian at a given order of $\lambda$. We will consider two examples corresponding to a truncation at order $\lambda^2$ and $\lambda^4$. If the Lagrangian is expanded to $\lambda^2$-order, we obtain for $D=4$:
		\begin{align}
			L^{(2)} & = e \wedge   h_{(1)} \wedge R(\omega) + \frac12  e \wedge e \wedge D(\omega) f_{(1)}  -\frac14 \mu^2 e \wedge e \wedge h_{(1)} \wedge h_{(1)} \,,
		\end{align}
where again we have refrained from giving  the explicit Lorentz indices.
		The field equation for the Lagrange multiplier $f_{(1)}$ implies that $\omega$ is a torsion-free spin connection while the $h_{(1)}$-equation determines $h_{(1)}$ to be the four-dimensional Schouten tensor. Upon substituting  the expressions for the spin-connection $\omega$ and the auxiliary field $h_{(1)}$  back into  the Lagrangian we obtain the following curvature squared Lagrangian:
		\begin{align}
			L^{(2)} & = - \frac{3}{4 \mu^2} \left( R_{\mu\nu} R^{\mu\nu} - \frac13 R^2 \right) \,.
		\end{align}
		The  dynamical equation for the Vierbein follows from this Lagrangian and, in contrast to the auxiliary field equations,  is not given by any truncation of the group-theoretical curvatures corresponding to  the infinite-dimensional algebra \eqref{4dExpandedAlgebra}.

		In $D$-dimensions, the same procedure yields
		\begin{align} \label{L2D}
			L^{(2)}_D & \sim  \frac{1}{\mu^2} \left( S_{\mu\nu} S^{\mu\nu} -  S^2 \right) \sim  \text{Gauss-Bonnet  -- (Weyl tensor)}^2\,,
		\end{align}
		where $S_{\mu\nu}$ is the $D$-dimensional Schouten tensor \eqref{SchoutenCotton}. For $D=3$, the combination \eqref{L2D} is precisely the NMG combination \cite{Bergshoeff:2009hq}. For $D \geq 4$,  the combination \eqref{L2D}  is precisely the $D$-dimensional term that is favored by the holographic c-theorem \cite{Liu:2010xc}. This can be seen from the fact that the  Gauss-Bonnet combination is the $D$-dimensional curvature-squared Lovelock theory that has second-order field equations admitting a holographic c-theorem whereas the second
		(Weyl tensor)$^2$ term does not contribute to the  holographic c-function.

	Finally, as a second example, we  consider a truncation at  $\lambda^4$-order. This leads to the following Lagrangian:
		\begin{align}
			L^{(4)} & = e \wedge h_2 \wedge R(\omega) + \frac12 h_1 \wedge h_1 \wedge R(\omega)  + e \wedge h_1 \wedge D(\omega)f _1 + \frac12 e \wedge e \wedge D(\omega) f_2  \nonumber\\
			& + \frac12 e \wedge e \wedge f_1 \wedge f_1 - \frac12 \mu^2 e \wedge e \wedge h_1 \wedge h_2   - \frac16 \mu^2 e \wedge h_1 \wedge h_1 \wedge h_1 \,.
		\end{align}
		Once again, one may verify that the set of soluble field equations are consistent with \eqref{4dExpandedAlgebra}, i.e.~the field equations for $f_2$ determine $\omega$ to be the torsionless spin connection while the field equations for $h_2$ and $f_1$ determine $h_1$ to be the Schouten tensor and $f_1$ to be the Cotton tensor, respectively. Substituting  these results back into the fourth-order Lagrangian, we obtain
		\begin{align}
			L^{(4)} & \sim \frac{1}{\mu^4} \left( R_{\mu\nu} R^{\nu\rho} R_{\rho}{}^\mu - R R^{\mu\nu} R_{\mu\nu} + \frac{7}{36} R^3 \right)  + \text{(Weyl and Cotton terms)}\,.
		\end{align}
		The $D$-dimensional generalization of this fourth-order Lagrangian  is given by
		\begin{align}
			L^{(4)}_D &  \sim \frac{1}{\mu^4} \left( S^3   - 3 S S^{\mu\nu} S_{\mu\nu}  + 2 S_{\mu\nu} S^{\nu\rho} S_{\rho}{}^\mu \right)  + \text{(Weyl and Cotton terms)}\,.
		\end{align}
		Like in the previous case, the first combination of terms between brackets  is precisely the $D$-dimensional (including $D = 3$ \cite{Sinha:2010ai,Paulos:2010ke}) fourth-order contribution that is favored by the holographic c-theorem \cite{Liu:2010xc}.
		In contrast, the second set of terms between brackets, containing the  Weyl and Cotton tensor, does not contribute to the  the holographic c-function.
		
		\section{Conclusions}
		\paragraph{}
		In this work we have discussed two types of infinite-dimensional Lie algebras: one, given in eq.~\eqref{4dExpandedAlgebra} for general $D$ and
		eq.~\eqref{Algebra} for 3D, that, upon truncation, loses its Lie algebra structure and gives the flavor tensors leading to higher-derivative gravity models that are polynomial in the curvatures and that are consistent with the holographic c-theorem and that in 3D describe extended massive gravity models. The other algebra, given in eq.~\eqref{CoAdjoint} for 3D, is identified as a Lie algebra expanded AdS$_D$ algebra, that allows consistent truncations describing a set of 1-forms coupled to gravity. The two algebras are closely related as indicated in eqs.~\eqref{shift} and \eqref{relation}. We have used the second AdS$_D$ algebra as a guide to come up with the first algebra \eqref{4dExpandedAlgebra} that underlies the holographic c-theorem.
		
		The work presented here serves as a starting point for various further studies. For instance, it would be much desired to understand the precise mathematical relationship between the two types of infinite-dimensional algebras \eqref{Algebra} and \eqref{CoAdjoint} beyond the shift given in eq.~\eqref{shift} and to see whether there is a  physical interpretation of this relationship such as some kind of symmetry breaking. The procedure for constructing special higher-derivative theories of gravity consistent with the holographic c-theorem can also be applied to other situations such as non-relativistic gravity models \cite{Bergshoeff:2019ctr} and, for 3D, to condensed matter physics \cite{Salgado-Rebolledo:2021wtf} where the Chern-Simons formulation of gravity plays an essential role. It may also be useful to realize the infinite-dimensional algebra and its truncation at the level of Hamiltonian and Poisson brackets as this may provide an insight on the degrees of freedom in higher-derivative gravity. Finally, there is a class of 3D gravity models that do not have a corresponding metric formulation with a single metric. These models are referred to as the third-way consistent models and they play an essential role in the resolution of the `bulk vs boundary clash in three dimensions \cite{Bergshoeff:2014pca,Bergshoeff:2015zga,Ozkan:2018cxj,Alkac:2018eck,Afshar:2019npk}. Our procedure does not include these models and it would be very interesting to see if there is an algebra that underlies the third way to gravity models. This might be the key to generalizing the third-way gravity models to higher dimensions which have been thus far an open problem.\,\footnote{For some recent results on third-way formulations without gravity in higher dimensions, see \cite{Broccoli:2021pvv}.}

		\medskip
		
		\noindent {\bf Acknowledgements}
		We thank G. Alkac,  D.O. Devecioglu, Y. Pang, P.K. Townsend, and U. Zorba for stimulating discussions. M.O. acknowledges the support by the Distinguished Young Scientist Award BAGEP of the Science Academy. M.O. also acknowledges the support by the Outstanding Young Scientist Award of the Turkish Academy of Sciences (TUBA-GEBIP).

		
		
	\end{document}